# WGM microprobe device for high-sensitivity and broadband ultrasound detection

Jialve Sun[1], Shengnan Huangfu[1], Tinglan Chen[2], Zijing Cai[2], Bowen Ruan[1], Fangxing Zhang[1]

[1] Peking University Yangtze Delta Institute of Optoelectronics, Nantong, Jiangsu 226019, China.

[2] School of Physics, Peking University, Beijing, 100871, China.

**Abstract:** Whispering-gallery-mode (WGM) microcavities have emerged as a promising alternative to traditional ultrasound probes, offering high sensitivity and wide bandwidth. In our research, we propose a novel silica WGM microprobe device, with impressive Q factors up to $10^7$. The side-coupled approach and special encapsulation design make the device small, robust, and capable of utilizing in both gaseous and liquid environments. We have successfully conducted photoacoustic (PA) imaging on various samples using this device which demonstrates a high sensitivity of 5.4 mPa/√Hz and a board bandwidth of 41 MHz at -6 dB for ultrasound. What's more, it's capable of capturing the vibration spectrum of microparticles up to a few hundred megahertz. Our compact and lightweight device exhibits significant application potential in PA endoscopic detection, near-field ultrasound sensing and other aspects.

## 1. Introduction

Ultrasound sensing finds widespread application across various domains, encompassing clinical ultrasound (US) imaging[1, 2], PA imaging[3, 4], and industrial non-destructive testing[5, 6], with the performance of the ultrasound detector being of paramount importance. In recent years, optical approaches for ultrasound detection, utilizing components such as microrings[7-9], Fabry–Pérot (FP) interferometers[10-12], and π-phase-shifted Bragg grating[13], have proved to outperform traditional piezoelectric-based ultrasound detectors in terms of both sensitivity and bandwidth[14]. Among those optical methods, owing to its substantially enhancing the interaction between light and ultrasound, the WGM-based US detectors have showcased remarkable progress and successful application in both cellular and in vivo PA imaging [15, 16].

Although the WGM microcavity has superiority in its diminutive size of tens of micrometers, the practical dimensions of the reported WGM microcavity US detectors (including coupling fiber and supporting substrate) usually reaches several millimeters or even larger at the sensing probe head[13, 16, 17], posing challenges to either implementations within very tight space (such as vascular endoscopy) or performing near-field-like detection over targets with uneven surface. Therefore, a needle-like miniature ultrasound probe based on sensitive WGM microcavity is extremely desired[18]. As a highly sensitive platform, microcavities are also expected to be able to use beyond the laboratory with more subtle measurements. In addition, due to the large frequency span (MHz-GHz) and the ultrahigh sensitivity requirements of mesoscopic vibration spectroscopy, microcavity devices are highly valued as the most effective solution, albeit very challengingly.

In this work, we propose a side-coupled method of $SiO_2$ microsphere to the folded fiber cone with controllable and robust encapsulation approach. Utilizing the spreadability and molecular surface tension of polymer microdroplets on the surface of silica microsphere, the tapered fiber is bound to the surface of the sphere for stable coupling with high Q factors. The sensor exhibits enhanced durability after fully encapsulation and the protection of the glass and metal tube. This innovative microprobe device boasts heightened ultrasound sensitivity (two orders of magnitude higher than piezoelectric hydrophones) and broad bandwidth, enabling the detection and imaging of high-frequency PA signals in liquid environments. It is worth mentioning that this device is also capable of capturing the the vibrational spectrum of mesoscale particles up to hundreds of megahertz by contact. The microprobe device has the ability to be used in various complex scenarios beyond the laboratory while

## 2. Experiment and Discussions

### 2.1 Structure design and working principle

Here we propose a side-coupled method for the fiber-based microsphere, incorporating a folded tapered fiber as shown in Fig 1. The microsphere cavity is coupled to the tip of the folded tapered fiber which send the pump light into the cavity and excite optical modes, as illustrated in the inset of Fig.1. The fiber aligns with the microsphere stem direction, ensuring a compact overall structure. We employ a thin layer of low refractive index polymer droplets to encapsulate the coupling area. To further enhance the device's durability, we reinforce it with a double-layer tube shell made of glass and aluminum alloy. In PA detection, the sample absorbs pulsed light and emits PA signal, which modulates the shape and refractive index of the microcavity. As a result, the microprobe device can effectively encode the ultrasound signal onto the optical mode with high sensitivity, enabling the extraction of amplitude and frequency information from the ultrasound signal through demodulation of the optical transmission signal.

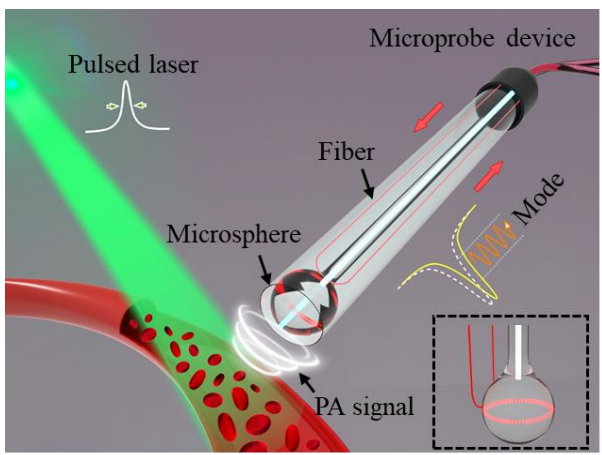

Fig.1 Structure of the microprobe device. It shows the coupling schematic (inset) and working principle of the microcavity mode for ultrasound signals.

### 2.2 Experimental results and performance measurements

Based on the proposed design, we successfully fabricated the microprobe device as shown in Fig.2(a). The microsphere sensing unit is located at the top of the tube shell. Standardized FC/APC port is implemented to enable plug-and-play utilization. The tube of aluminum alloy is just 2 mm. In Fig.2(b), the encapsulated structure of the sensing unit is displayed, with the microsphere measuring approximately 70 μm in size. The tapered fiber (diameter ~1 μm) is attached to the stem of the microsphere, ensuring an entire structure integrated. Characterization of the microcavity modes was conducted using a tunable laser (TOPICA, 910-980 nm), with the results presented in Fig. 2(c). The analysis in Fig. 2(d) reveals an impressive maximum Q factor of $2\times10^7$ for the microcavity. The compact dimensions and high Q of the microprobe facilitate an extremely high energy density of light for high sensitivity sensing. Furthermore, the integrated coupling structure enables versatile utilization of the device in both air and liquid environments.

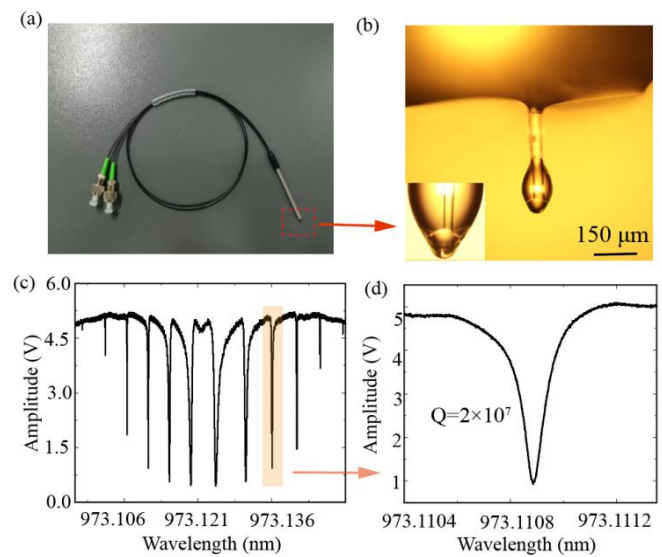

Fig.2 (a) Photograph of the microprobe device after packaging and ruggedization. (b) Enlarged view of the microprobe sensing unit. (c) The modes distribution of the microsphere tested by a tunable laser. (d) Enlarged view of one mode in (c).

Further, we tested the ultrasound response of the microprobe device with results shown in Fig.3. The sensitivity of our microprobe device is measured by detecting 20 MHz ultrasound waves emitted from a calibrated ultrasound transducer driven by a pulse/echo receiver. The detected signal after a 1.2–800 MHz band-pass filter is illustrated in Fig. 3(a). Notably, the noise equivalent pressure (NEP) is calculated to be as low as 24 Pa, or equally 5.4 mPa√Hz. This heightened sensitivity can be attributed to the exceptional Q factor of the microcavity. At the same time, the diminutive size of the cavity results in a significantly high energy density of pump light, which is also of benefit to sensitivity. To evaluate the frequency response of the device, we used a pulsed laser (pulse width ~1.8 ns, pulse energy ~2 μJ) to irradiate a copper film with a thickness of 50 nm to generate a broadband PA signal. The detected time-domain PA signal is shown in Fig. 3(b), and its Fourier transform outcome is shown in Fig. 3(c). The results indicate a broadband response exceeding 40 MHz in -6 dB. This solution of integrating the cavity and the stem can make the ultrasound pass through the microcavity and be guided away, thus mitigating the occurrence of multiple sound waves reflections within the cavity. We also calibrated the response of the microprobe to ultrasound from different angles as shown in Fig.3(d). The response is within -6 dB over a range of 180 degrees. As a point detector, the microsphere cavity can theoretically respond uniformly to the full space angle. However, the relative position of the protective shell to the microsphere may block some of the ultrasound. By adjusting the shell position, the receiving angle can be tailored as required.

## 2.3 Photoacoustic imaging with microprobe device

We built the PA microscopy (PAM) imaging system which is basically the same as the system in our previous study[17, 18]. The pulsed light is focused by the objective lens and irradiates the sample from below. The microcavity probe captures the PA signal from above the sample, and the translation stage drives the sample for 2D scanning. During imaging, both the sample and microprobe are immersed in an aqueous environment. PAM was conducted on different samples, yielding compelling results. Fig.4(a) shows the PA image of hairs that are tightened and fixed to the glass slide. Not only the contours of the hair can be delineated in the PA image, but also the scaly structure on the surface of the hair. Fig.4(b) shows two curly hairs, buried in a hydrogel for PA imaging, still achieving high contrast. In addition, we performed PAM of a gold film engraved with letters in Fig.4(c). The width of the font is around 50 μm, displaying a clear outline edge. In Fig.4(d), We performed PAM of an ant that was immobilized in a hydrogel, showcasing a well-defined silhouette of the ant. However, one leg is not clearly displayed because it is bent and not in the focus of the pulsed light. These PAM examples demonstrate the exceptional performance and stability of our microprobe device. Each image takes no more than 15 minutes, and the optimal resolution of the image can reach ~2 μm.

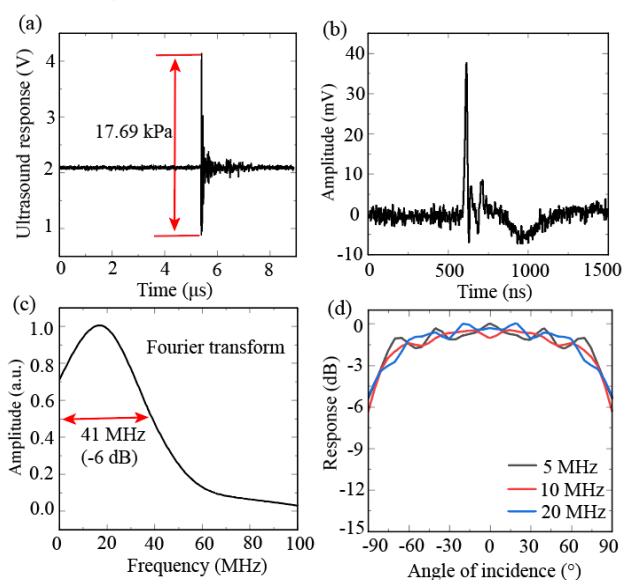

Fig.3 (a) The response of the microprobe device to the emitting signal of an ultrasound transducer. (b) The response of the microprobe device to a broadband PA signal. (c) Fourier transform of the photoacoustic response in (b). (d) The response of the microprobe to the ultrasound at different incident angles.

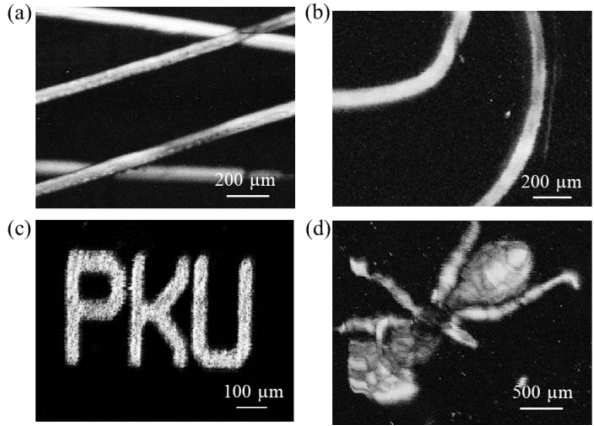

Fig.4 PA imaging of samples. (a) Tight and crossed hairs. (b) Curly hairs. (c) Letter-shaped gold film. (d) Ant.

**2.4 Mesoscopic vibrational spectroscopy**

Vibrational spectroscopy is a ubiquitous technology that derives the species, constituents and morphology of an object from its natural vibrations. Despite the challenges associated with the measurement of mesoscopic vibration spectroscopy, it presents new opportunities for the elucidation of various entities, including all kinds of particles[19, 20], as well as biological cells and viruses[21, 22]. Microcavity-based detector to conduct vibrational spectroscopy measurements of different bacteria and micro-nanoparticles is feasible in laboratory cleanrooms [23]. Here, our microprobe device can not only be suitable for micro/nano particles' broadband vibration spectroscopy measurements, but also hold significant promise for application and promotion beyond the laboratory. The microsphere is surrounded by a layer of polymer glue of a thickness of several microns, which will not impede the transmission of high-frequency ultrasound while fixing the coupling point. By focusing the 532 nm pulsed light on the particles adhering to the microprobe, the resultant broadband PA signal encompasses the entire vibrational spectrum of the particles. The signal of the mechanical spectrum is enhanced by resonance, and other frequencies are rapidly attenuated. The detected electrical signals were amplified (SHF 806E, 26 dB at 40 kHz - 38 GHz) and recorded by an oscilloscope (Keysight, DSOS254A) with 64 times average. The vibrational spectroscopy measurements of polystyrene particles with a radius about 2.8 μm are shown in Fig.5. The diagram illustrates the time-domain signal of vibrations in (a) and its Fourier spectrum distribution is shown in (b). This particle exhibits vibrations exceeding 100 ns, with the main vibration peak located at 355 MHz. The results are in agreement with the theory[23] given the material density of polystyrene is 1.05 g/cm$^{-3}$, Young's modulus is 3.24 GPa, and Poisson ratio is 0.34. This method provides a novel avenue for the measurement of mesoscopic particle vibrational spectroscopy. It is essential to note that bandwidth measurement in PA imaging entails detecting of pulsed signals, while the vibrational spectroscopy involves continuous acoustic sensing. Therefore, the bandwidth of PA in -6 dB is different from the measurement range of the vibration spectrum. With sufficient sensitivity, the microcavity can measure vibrations well below -6 dB.

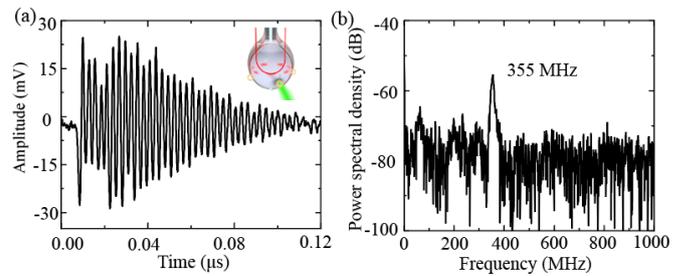

Fig.5 (a) Time-domain vibration signal from the polystyrene particles excited by pulsed light. The inset shows the schematic of measurement. (b) The vibrational spectrum in the frequency domain corresponding to (a).

**3. Conclusions**

In this article, we have successfully demonstrated a WGM microprobe device with high Q factors. A novel coupling and encapsulation approach is proposed for ultrasensitive ultrasound detection and photoacoustic imaging. The coupling region is securely enclosed by a thin layer of low refractive index polymer, ensuring robust stability of the microcavity mode without compromising the detection of ultrasound. Our comprehensive evaluation encompassed the ultrasound response, the successful execution of photoacoustic imaging on diverse samples, and the measurement of vibration spectrum of mesoscopic objects with microcavity probes.

While we have highlighted two typical applications, the versatility of this microprobe can extend to various scenarios such as PA endoscopic imaging, gas/liquid PA spectroscopy, and metal flaw detection. Compared to microcavities on chips, our structure features a simpler package design, greater versatility in applications, and notable advantages in ultrasound response performance [24, 25]. This coupling structure offers considerable flexibility. By adjusting the size of the microsphere cavity, it can be applied to ultrasound measurements with different bandwidths and different center frequencies. It is applicable for ultrasound measurement in air as well as for detecting ultrasound signals in aqueous or other liquid environments. And from the preparation of microcavities to packaging and deviceization, the whole process is controllable, and the success rate is over 80%. We firmly believe that the microprobe design proposed in this paper holds significant potential for applications in ultrasound detection and vibration measurement, promising advancements in various fields of high-sensitivity ultrasound sensing.

## Declarations

**Availability of data and material**
The data and materials used in this study are available from the corresponding author upon reasonable request.

**Funding**
This work was supported by the National Natural Science Foundation of China (Grant No.62305006), Natural Science Foundation of Jiangsu Province (Grant Nos. BK20230287, BK20230286), Nantong Social Livelihood Science and Technology Planning Project (Grant Nos. MS12022003, MS2023071).
**Acknowledgements**
We acknowledge S.-J.T for the help with the measurement of vibrational spectroscopy.
**Disclosures**
The authors declare no conflicts of interest.

## References

1. J. Powers, and F. Kremkau, "Medical ultrasound systems," Interface Focus **1**, 477-489 (2011).
2. W. P. N. T., "Biomedical Ultrasonics.," Academic Press (1977).
3. L. V. Wang, and J. Yao, "A practical guide to photoacoustic tomography in the life sciences," Nat Methods **13**, 627-638 (2016).
4. L. H. V. Wang, and S. Hu, "Photoacoustic Tomography: In Vivo Imaging from Organelles to Organs," Science **335**, 1458-1462 (2012).
5. M. Shaloo, M. Schnall, T. Klein, N. Huber, and B. Reitinger, "A Review of Non-Destructive Testing (NDT) Techniques for Defect Detection: Application to Fusion Welding and Future Wire Arc Additive Manufacturing Processes," Materials (Basel) **15** (2022).
6. B. W. Drinkwater, and P. D. Wilcox, "Ultrasonic arrays for non-destructive evaluation: A review," Ndt&E Int **39**, 525-541 (2006).
7. H. Li, B. Dong, Z. Zhang, H. F. Zhang, and C. Sun, "A transparent broadband ultrasonic detector based on an optical micro-ring resonator for photoacoustic microscopy," Sci Rep **4**, 4496 (2014).
8. C. Zhang, T. Ling, S.-L. Chen, and L. J. Guo, "Ultrabroad Bandwidth and Highly Sensitive Optical Ultrasonic Detector for Photoacoustic Imaging," ACS Photonics **1**, 1093-1098 (2014).
9. W. J. Westerveld, Mahmud-Ul-Hasan, Md, Shnaiderman, Rami, Ntziachristos, Vasilis, Rottenberg, Xavier, Severi, Simone, Rochus, Veronique, "Sensitive, small, broadband and scalable optomechanical ultrasound sensor in silicon photonics," Nature Photonics **15**, 341-345 (2021).
10. E. Zhang, J. Laufer, and P. Beard, "Backward-mode multiwavelength photoacoustic scanner using a planar Fabry-Perot polymer film ultrasound sensor for high-resolution three-dimensional imaging of biological tissues," Appl Opt **47**, 561-577 (2008).
11. J. A. Guggenheim, Li, Jing, Allen, Thomas J., Colchester, Richard J., Noimark, Sacha, Ogunlade, Olumide, Parkin, Ivan P., Papakonstantinou, Ioannis, Desjardins, Adrien E., Zhang, Edward Z., Beard, Paul C., "Ultrasensitive plano-concave optical microresonators for ultrasound sensing," Nature Photonics **11**, 714-719 (2017).
12. A. P. Jathoul, Laufer, Jan, Ogunlade, Olumide, Treeby, Bradley, Cox, Ben, Zhang, Edward, Johnson, Peter, Pizzey, Arnold R., Philip, Brian, Marafioti, Teresa, Lythgoe, Mark

F., Pedley, R. Barbara, Pule, Martin A., Beard, Paul, "Deep in vivo photoacoustic imaging of mammalian tissues using a tyrosinase-based genetic reporter," Nature Photonics **9**, 239-246 (2015).

13. Y. Hazan, A. Levi, M. Nagli, and A. Rosenthal, "Silicon-photonics acoustic detector for optoacoustic micro-tomography," Nat Commun **13**, 1488 (2022).

14. G. Wissmeyer, Pleitez, M. A., Rosenthal, A., Ntziachristos, V., "Looking at sound: optoacoustics with all-optical ultrasound detection," Light Sci Appl **7**, 53 (2018).

15. C. Zhang, S.-L. Chen, T. Ling, and L. J. Guo, "Imprinted Polymer Microrings as High-Performance Ultrasound Detectors in Photoacoustic Imaging," Journal of Lightwave Technology **33**, 4318-4328 (2015).

16. H. Li, Dong, B., Zhang, X., Shu, X., Chen, X., Hai, R., Czaplewski, D. A., Zhang, H. F.,Sun, C., "Disposable ultrasound-sensing chronic cranial window by soft nanoimprinting lithography," Nat Commun **10**, 4277 (2019).

17. J. Sun, Meng, Jia-Wei, Tang, Shui-Jing, Li, Changhui, "An encapsulated optical microsphere sensor for ultrasound detection and photoacoustic imaging," Science China Physics, Mechanics & Astronomy **65** (2021).

18. J. Sun, Tang, Shui-Jing, Meng, Jia-Wei, Li, Changhui, "Whispering-gallery optical microprobe for photoacoustic imaging," Photonics Research **11**, A65 (2023).

19. M. Tayebi, R. O'Rorke, H. C. Wong, H. Y. Low, J. Han, D. J. Collins, and Y. Ai, "Massively Multiplexed Submicron Particle Patterning in Acoustically Driven Oscillating Nanocavities," Small **16** (2020).

20. H. Keshtgar, S. Streib, A. Kamra, Y. M. Blanter, and G. E. W. Bauer, "Magnetomechanical coupling and ferromagnetic resonance in magnetic nanoparticles," Physical Review B **95** (2017).

21. P. V. Zinin, J. S. Allen, and V. M. Levin, "Mechanical resonances of bacteria cells," Phys Rev E **72** (2005).

22. E. C. Dykeman, and O. F. Sankey, "Low Frequency Mechanical Modes of Viral Capsids: An Atomistic Approach," Physical Review Letters **100** (2008).

23. S.-J. Tang, M. Zhang, J. Sun, J.-W. Meng, X. Xiong, Q. Gong, D. Jin, Q.-F. Yang, and Y.-F. Xiao, "Single-particle photoacoustic vibrational spectroscopy using optical microresonators," Nature Photonics (2023).

24. J. Sun, F. Hou, S. Feng, and C. Li, "Integrated Optical Microrings on Fiber Facet for Broadband Ultrasound Detection," Advanced Sensor Research (2024).

25. Z. Ding, J. Sun, C. Li, and Y. Shi, "Broadband Ultrasound Detection Using Silicon Micro-Ring Resonators," Journal of Lightwave Technology **41**, 1906-1910 (2023).